\documentclass[a4paper,11pt]{article}
\pdfoutput=1 % if your are submitting a pdflatex (i.e. if you have
\usepackage{jinstpub} % for details on the use of the package, please
\usepackage[utf8]{inputenc}
\usepackage{circuitikz}
\usepackage{multirow}
\usepackage{booktabs}
\usepackage{comment}
\usepackage{float}

\title{\boldmath Radiation Damage and Recovery Properties of Common Plastics PEN (Polyethylene Naphthalate) and PET (Polyethylene Terephthalate) Using a $^{137}$Cs Gamma Ray Source Up To 1 MRad and 10 MRad}

\author[a,1]{J. Wetzel,\note{Corresponding author.}}
\author[a]{E. Tiras,}
\author[a]{B. Bilki,}
\author[a]{Y. Onel,}
\author[b]{D. Winn}

\affiliation[a]{Department of Physics and Astronomy, University of Iowa\\203 Van Allen Hall, Iowa City, IA 52242, USA}
\affiliation[b]{Department of Physics, Fairfield University\\BNW118, 1073 N. Benson Rd., Fairfield, CT 06824, USA }

\emailAdd{james-wetzel@uiowa.edu}

\abstract{Polyethylene naphthalate (PEN) and polyethylene teraphthalate (PET) are cheap and common polyester plastics used throughout the world in the manufacturing of bottled drinks, containers for foodstuffs, and fibers used in clothing. These plastics are also known organic scintillators with very good scintillation properties.  As particle physics experiments increase in energy and particle flux density, so does radiation exposure to detector materials.  It is therefore important that scintillators be tested for radiation tolerance at these generally unheard of doses.  We tested samples of PEN and PET using laser stimulated emission on separate tiles exposed to 1 MRad and 10 MRad gamma rays with a $^{137}$Cs source.  PEN exposed to 1 MRad and 10 MRad emit 71.4\% and 46.7\% of the light of an undamaged tile, respectively, and maximally recover to 85.9\% and 79.5\% after 5 and 9 days, respectively.  PET exposed to 1 MRad and 10 MRad emit 35.0\% and 12.2\% light, respectively, and maximally recover to 93.5\% and 80.0\% after 22 and 60 days, respectively.
}

\keywords{Scintillators, scintillation and light emission processes (solid, gas and liquid scintillators), Radiation-hard detectors, Radiation damage to detector materials (solid state)}

\begin{document}
	\maketitle
	\flushbottom

	\section{Introduction}
		\label{sec:intro}
		Polyethylene naphthalate (PEN) and polyethylene teraphthalate (PET) are known organic scintillators with peak emission wavelengths of ~450 nm and ~350 nm, respectively \cite{Nakamura}.  These are attractive additions to the arsenal of high energy physics experiments' organic scintillators because of their ease of manufacture and low cost.  However, high energy physics experiments like those around the LHC \cite{evans2008lhc} and future colliders \cite{benedikt2014future} are creating increasingly problematic radiation zones for the detector equipment, therefore, before considering PEN or PET to replace existing scintillator technologies, it is important to understand how they respond to high levels of radiation.
		
		We irradiated two sets of PEN and PET tiles to 1 MRad and 10 MRad doses of gamma radiation under a long rod of $^{137}$Cs at the University of Iowa RadCore facility \cite{radcore}, and measured their emittance levels before irradiation and after irradiation, over more than 60 days.  Our goal was to understand the emittance attenuation for low dose vs. high dose exposure to PEN and PET, as well as the subsequent recovery time of each sample at each dose.
		
		We used 100 mm x 100 mm x 2 mm sheets of PET, and 100 mm x 100 mm x 1 mm sheets of PEN.  The PET sheets were purchased from Goodfellow.com \cite{goodfellow}.  The PEN sheets were purchased from Teijin \cite{Teijin}, under the brand name Scinterex \cite{Scintirex}, their scintillator formulation of PEN.

	\section{Experimental Setup}

		Three tiles of each scintillator were cut to 100 x 100 mm squares, and the edges were polished.  Each tile was tested prior to radiation damage and each tile had measurement results within 5\% of each other, showing consistency between tiles.  One tile was kept in the dark, not irradiated, for reference, and the other two were placed on the lucite tray used to install the tiles into the radiation source.  One sample of the remaining two was placed on the top shelf, the other on the bottom shelf.  The top shelf sits closer to the source, and therefore receives a 10x higher dose rate.
		
		The lucite tray was then mounted against the source, and irradiated for 67.614 hours, for a total of 10.1421 MRad on the high shelf, and 1.0142 MRad on the low shelf.
		
		Each tile was tested within an hour after removal from the radiation source.  Then they were tested twice each day for the first 2 days after exposure, and then once per day for approximately 20 days, and then once per week for over 60 days after the initial exposure.
		
		Before every test, a reference sample of a generic scintillator was used to ensure stability of measurements in order to reduce systematics from laser jitter and PMT warm up effects.
		
		Between tests, each sample was kept in a light-tight briefcase in order to limit exposure to ambient light, which might influence recovery.

\begin{figure}[]
			\begin{center}
				\includegraphics[width=0.6\textwidth]{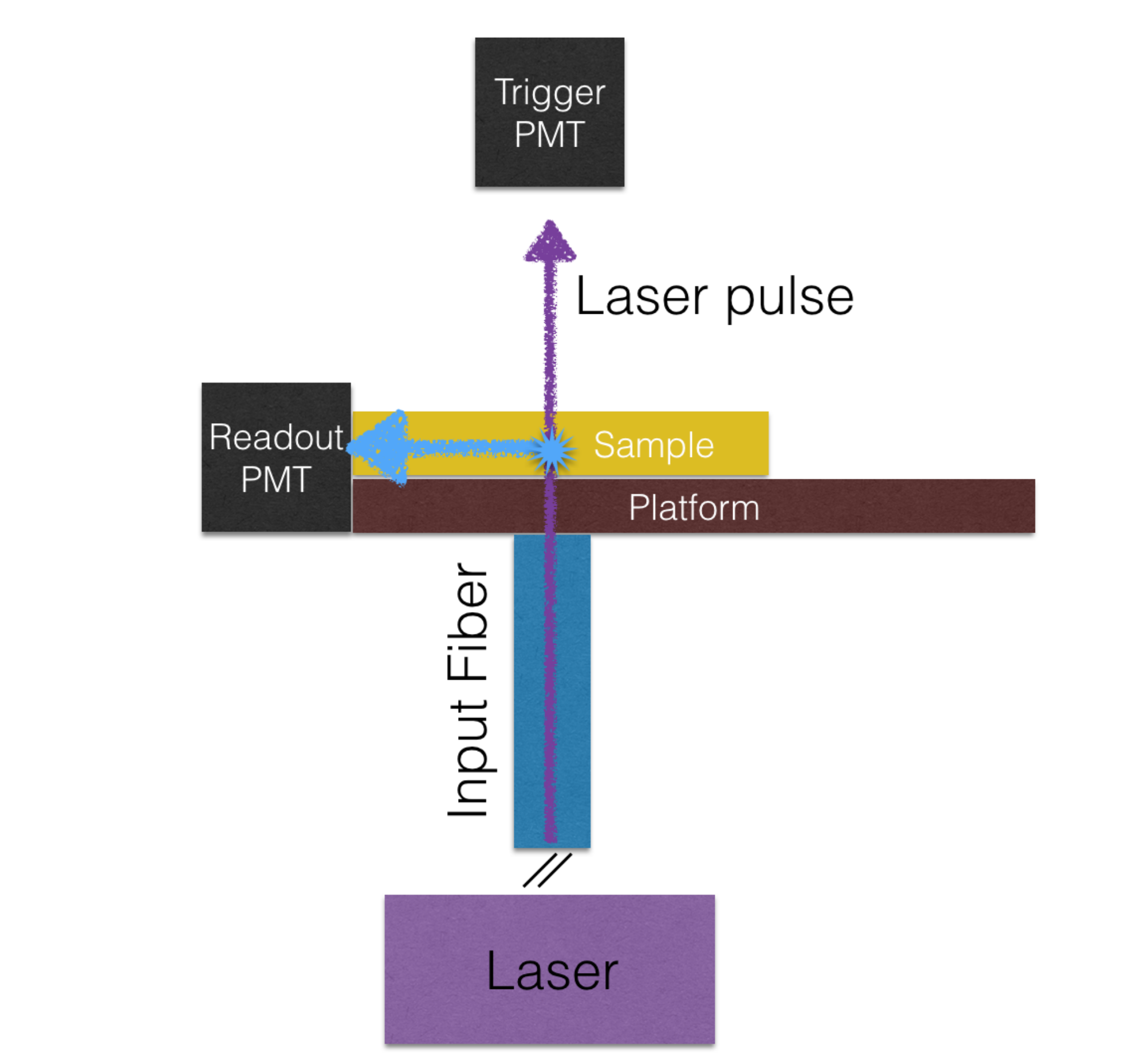}
				\caption{Experimental test setup.}
				\label{fig:LaserSetup}
			\end{center}
		\end{figure}

		The tiles were tested by pulsing a 3 ns pulse-width 337 nm nitrogen laser perpendicular to the tile surface, and reading out the scintillated light from the edge of the tile using a Hamamatsu R7600 PMT \cite{R7600}, Figure \ref{fig:LaserSetup}.  A separate Hamamatsu R7525 PMT \cite{R7525} was placed directly above the tile to be used as a trigger for the measurements.
		
		%%%%%%%%%%%%%%%%% Figure Laser Setup %%%%%%%%%%%%%%%
		%%%%%%%%%%%%%%%%%%%%%%%%%%%%%%%%%%%%%%%%%%

	\section{Measurement Results}
\begin{figure}[H]
			\begin{center}
				\includegraphics[width=0.41\textwidth]{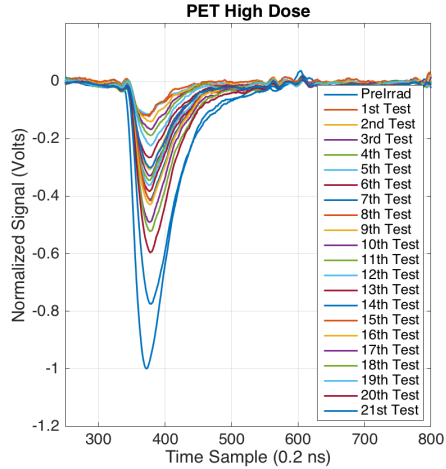}
				\caption{Average waveforms of each of 21 measurements over 40 days of the PET 10 MRad dosed sample.}
				\label{fig:PETWaveform}
			\end{center}
		\end{figure}

		We measured the light output from each tile before irradiation, and began testing within 1 hour of removal from the RadCore facility and continued until maximum recovery of the tile was observed.  Each measurement was an average of 100 laser pulses using a Tektronics TDS5034 oscilloscope \cite{TDS5034}.  Figure \ref{fig:PETWaveform} shows for an example of the average waveforms of each of 21 measurements over 40 days of the sample of PET irradiated to 10 MRad.
		
		Tables \ref{table:PENResults} and \ref{table:PETResults} summarize the measurement results for PEN and PET respectively.  The first column in the table labels the sample, 1 MRad or 10 MRad.  The `initial emittance' column describes the measured light output compared to the undamaged tile immediately after the exposure.  After removal from the radiation source, the tiles slowly increase their emittance and plateau to a maximum value.  We refer to this period as the recovery period, and refer to the maximum value as the `recovered emittance' in the third column of the tables.  The time it takes to reach the plateau is listed in column 4, as `recovery time'.
		
		%%%%%%%%%%%%%%%%% High Dose PET %%%%%%%%%%%%%%%%
		%%%%%%%%%%%%%%%%%%%%%%%%%%%%%%%%%%%%%%%%%%
		
		\begin{table}[H]
			\caption{Summary of PEN irradiation results}
			\label{table:PENResults}
			\begin{center}
				\begin{tabular}{lccc}
					\toprule
						{Total Dose} & {Initial Emittance} & {Recovered Emittance} & {Recovery Time (days)}\\
						\midrule
						{1 Mrad} & {71.4 ${\pm}$ 2.4\%} & {85.9 ${\pm}$ 2.4\%} & {5}\\
						{10 Mrad} & {46.7${\pm}$ 2.7\%} & {79.5 ${\pm}$ 2.7\%} & {9}\\
						\bottomrule
				\end{tabular}
			\end{center}
		\end{table}
		
		\begin{table}[H]
			\caption{Summary of PET irradiation results}
			\label{table:PETResults}
			\begin{center}
				\begin{tabular}{lccc}
					\toprule
						{Total Dose} & {Initial Emittance} & {Recovered Emittance} & {Recovery Time (days)}\\
						\midrule
						{1 Mrad} 	& {35.0 ${\pm}$ 2.4\%} & {93.5 ${\pm}$ 2.4\%} & {22}\\
						{10 Mrad} & {12.2${\pm}$ 2.7\%} & {80.0 ${\pm}$ 2.7\%} & {60}\\
						\bottomrule
				\end{tabular}
			\end{center}
		\end{table}
				
\begin{figure}[H]
			\begin{center}
				\includegraphics[width=1.0\textwidth]{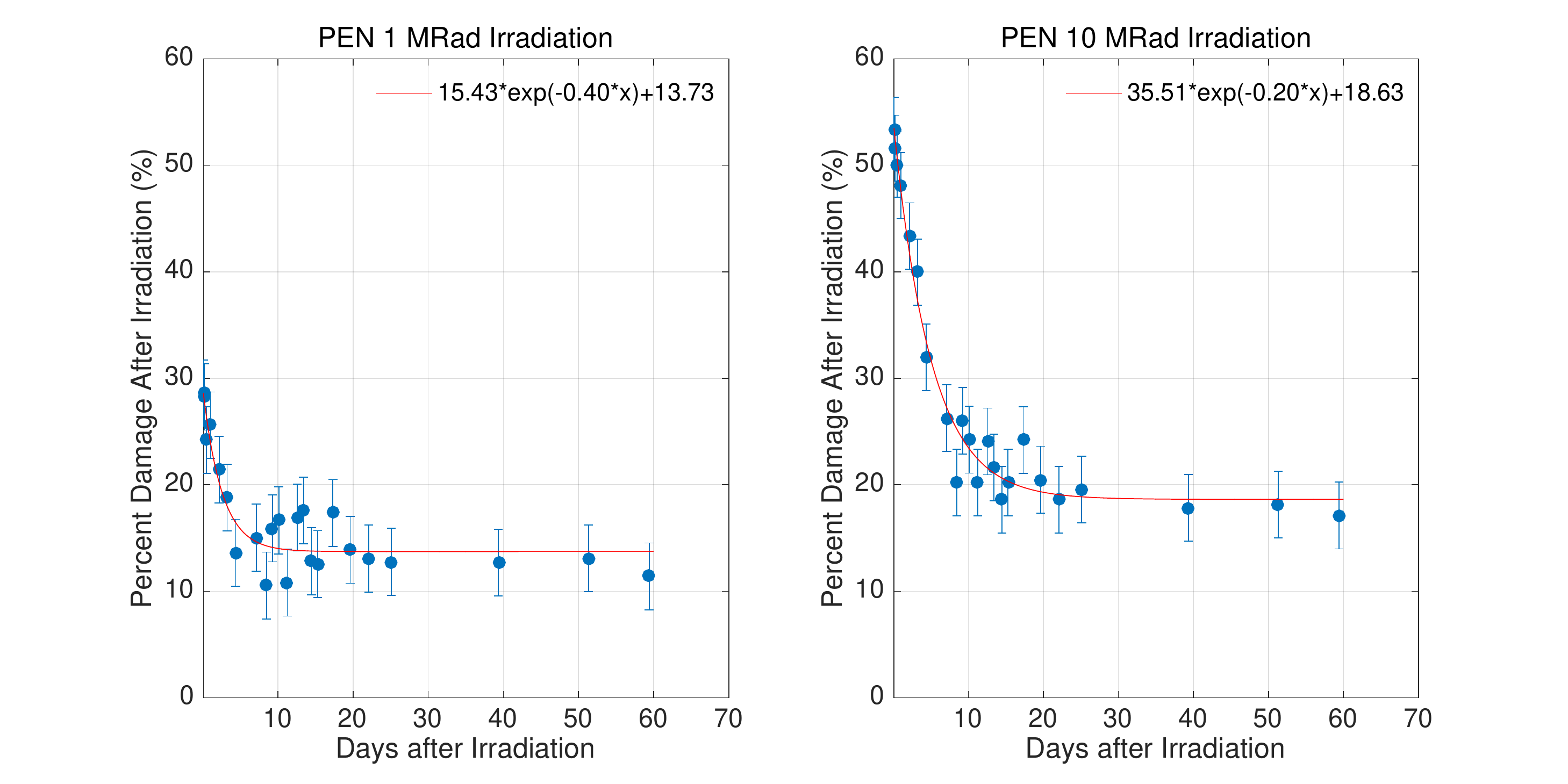}
				\caption{PEN Results, 21 measurements over 40 days.}
				\label{fig:PENResults}
			\end{center}
		\end{figure}

\begin{figure}[H]
			\begin{center}
				\includegraphics[width=1.0\textwidth]{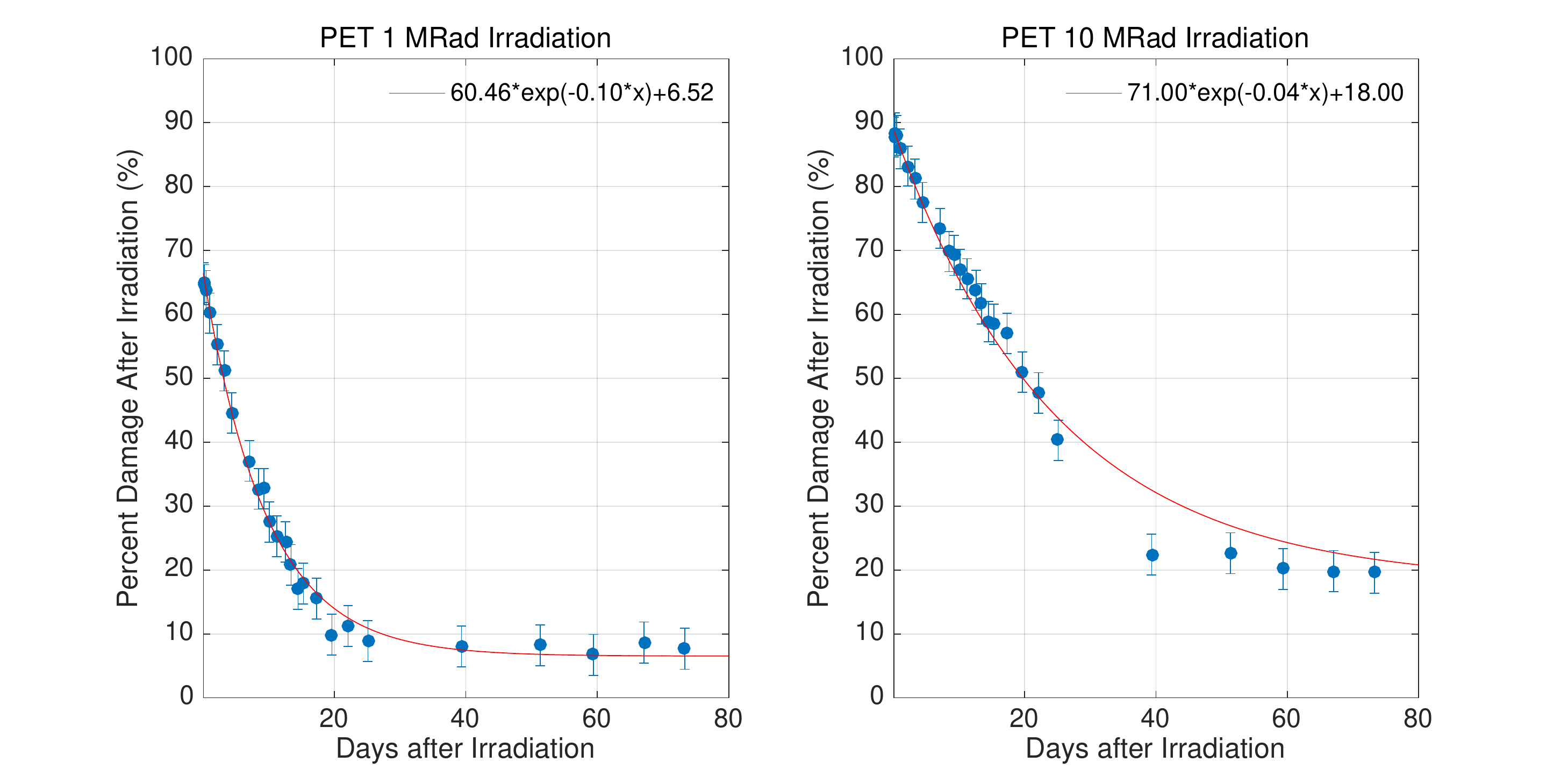}
				\caption{PET Results, 21 measurements over 40 days.}
				\label{fig:PETResults}
			\end{center}
		\end{figure}

		The 1 MRad and 10 MRad results can be seen in Figures \ref{fig:PENResults} and \ref{fig:PETResults}, left and right, for PEN and PET, respectively.  Each data point in Figures \ref{fig:PENResults} and \ref{fig:PETResults} is the integral of the averaged waveforms shown in Figure \ref{fig:PETWaveform}, subtracted from 100\%.  Repeated testing of the reference tile yielded a standard deviation of 3\%, which we have applied as an error to each measurement.

		%%%%%%%%%%%%%%%%% PEN Results %%%%%%%%%%%%%%%%%%
		%%%%%%%%%%%%%%%%%%%%%%%%%%%%%%%%%%%%%%%%%%
		
		%%%%%%%%%%%%%%%%% PET Results %%%%%%%%%%%%%%%%%%
		%%%%%%%%%%%%%%%%%%%%%%%%%%%%%%%%%%%%%%%%%%
		
		The plots include fits of the form `$a e^{-b x}+c$', where `$a+c$' gives the initial percent damage, `$c$' describes the permanent damage, and `$1/b$' characterizes the rate of recovery.
		
		%%%%%%%%%%%%%%%%% PEN/PET Dose vs Damage Results %%%%%%%%%%%%%%%%%%
		\begin{figure}[H]
			\begin{center}
				\includegraphics[width=1.0\textwidth]{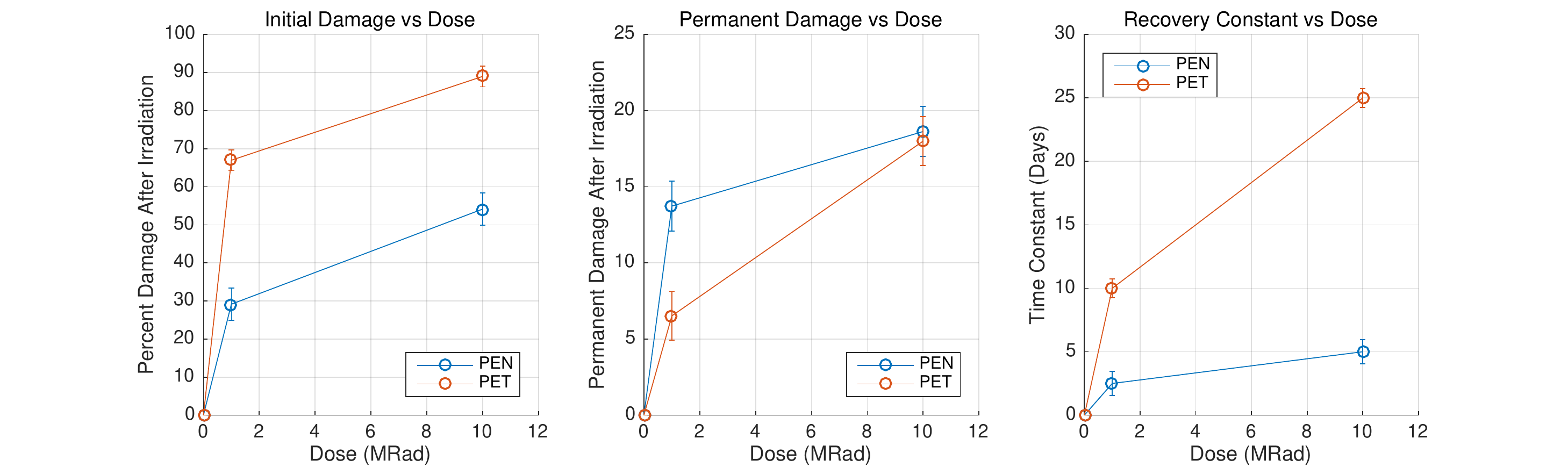}
				\caption{Damage vs Dose characteristics for PEN and PET, comparing total dose with initial damage, permanent damage, and recovery time.}
				\label{fig:DamVsDose}
			\end{center}
		\end{figure}
		%%%%%%%%%%%%%%%%%%%%%%%%%%%%%%%%%%%%%%%%%%%%%%%%%%%%%		
	
		Figure \ref{fig:DamVsDose} plots these values as a function of dose and compares PEN to PET.  The leftmost plot shows the initial damage immediately after irradiation.  The center plot compares the permanent damage between the tiles.  The rightmost graph plots the recovery constant.  These plots show that while PEN handles a high radiation environment better than PET, PET has a remarkable ability to recover.  PET, however, takes a considerable amount of time to recover.
		
		The damage also appears to be correlated with exposure dose following a logarithmic trend, consistent with Oldham and Ware \cite{oldham1975} and Buss et al. \cite{bussAndDannemann}.  In other words, most of the damage occurs early in the exposure, for example PET was damaged 65\% in the first 1 MRad of exposure, but increased only 23\% further after another 9 MRad.
	
	\section{Conclusions}
		Samples of Polyethylene Naphthalate (PEN) and Polyethylene Teraphthalate (PET) were dosed to two different amounts, 1 MRad and 10 MRad.  Both exhibited a greater reduction in emittance with higher dose as expected.  PET was significantly more sensitive to the radiation than PEN, by a factor of 2 for the 1 MRad dose, and a factor of 3.8 for the 10 MRad dose.  PEN also reached maximum recovery more quickly than PET by a significant margin.  The lower sensitivity to radiation, and the shorter recovery period indicate that PEN is a better candidate for consideration in high radiation zones.  A possible copolymer of PEN and PET might reveal superior/optimal radiation-hardness and recovery specifications when compared to the individual components.

	\acknowledgments
		The authors would like to thank Amanda Kalen for technical assistance at the Radiation and Free Radical Research core supported in part by the Holden Comprehensive Cancer Center Grant P30 CA086862 and the University of Iowa Carver College of Medicine.


\begin{thebibliography}{10}
	
	\bibitem{Nakamura}
	H.~Nakamura, Y.~Shirakawa, S.~Takahashi, and H.~Shimizu.
	\newblock Evidence of deep-blue photon emission at high efficiency by common
	  plastic.
	\newblock {\em EPL (Europhysics Letters)}, 95(2):22001, 2011.
	
	\bibitem{evans2008lhc}
	Lyndon Evans and Philip Bryant.
	\newblock Lhc machine.
	\newblock {\em Journal of Instrumentation}, 3(08):S08001, 2008.
	
	\bibitem{benedikt2014future}
	Michael Benedikt and F~Zimmermann.
	\newblock The future circular collider study.
	\newblock Technical report, 2014.
	
	\bibitem{radcore}
	{The University of Iowa Research Core}.
	\newblock \url{http://frrbp.medicine.uiowa.edu/research-core}.
	\newblock Accessed: 2016-4-27.
	
	\bibitem{goodfellow}
	Goodfellow.
	\newblock
	  \url{http://www.goodfellow.com/catalogue/GFCat4I.php?ewd_token=Z8SWQhScVbDVbDN1baPtqu4xkXWAyx&n=wbpTWomjct5vzVtdgHQL61B3udNxHH&ewd_urlNo=GFCatSeaReb6&Catite=ES303021&CatSearNum=2}.
	\newblock Accessed: 2016-4-27.
	
	\bibitem{Teijin}
	{Teijin Limited.}
	\newblock \url{http://www.teijin.com}.
	\newblock Accessed: 2016-4-27.
	
	\bibitem{Scintirex}
	{Scintirex Brand Scintillator.}
	\newblock \url{http://www.teijin.com/news/2011/ebd110907_00.html}.
	\newblock Accessed: 2016-4-27.
	
	\bibitem{R7600}
	{Hamamatsu Photonics}.
	\newblock
	  \url{http://www.hamamatsu.com/resources/pdf/etd/High_energy_PMT_TPMO0007E.pdf}.
	\newblock Accessed: 2016-4-27.
	
	\bibitem{R7525}
	U~Akgun, A~S Ayan, G~Aydin, F~Duru, J~Olson, and Y~Onel.
	\newblock Afterpulse timing and rate investigation of three different hamamatsu
	  photomultiplier tubes.
	\newblock {\em Journal of Instrumentation}, 3(01):T01001, 2008.
	
	\bibitem{TDS5034}
	{Tektronix TDS5034 Oscilloscope.}
	\newblock
	  \url{http://www.tek.com/datasheet/tds5000-series-digital-phosphor-oscilloscope}.
	\newblock Accessed: 2015-10-08.
	
	\bibitem{oldham1975}
	G.~Oldham and A.~R. Ware.
	\newblock Gamma-radiation damage effects on plastic scintillators.
	\newblock {\em Radiation Effects}, 26(1-2):95--97, 1975.
	
	\bibitem{bussAndDannemann}
	G.~Buss, A.~Dannemann, U.~Holm, and K.~Wick.
	\newblock Radiation damage by neutrons to plastic scintillators.
	\newblock {\em IEEE Transactions on Nuclear Science}, 42(4):315--319, Aug 1995.
	
	\end{thebibliography}
\end{document}